\documentstyle[12pt,epsfig]{article}
\newcommand{\beq}{\begin{equation}}
\newcommand{\eeq}{\end{equation}}
\newcommand{\bea}{\begin{eqnarray}}
\newcommand{\eea}{\end{eqnarray}}

\newcommand{\gsim}{\lower.7ex\hbox{$
\;\stackrel{\textstyle>}{\sim}\;$}}
\newcommand{\lsim}{\lower.7ex\hbox{$
\;\stackrel{\textstyle<}{\sim}\;$}}

\newcommand{\eod}{\end{document}}

\def\ot{{\bf T}}
\def\cp{{\bf CP}}

\begin{document}
\thispagestyle{empty}
\vspace*{-22mm}

\begin{flushright}
UND-HEP-08-BIG\hspace*{.08em}01\\
\today

\end{flushright}
\vspace*{1.3mm}

\begin{center}
{\LARGE{\bf
"Ceterum Censeo Fabricam Super Saporis Esse Faciendam" \\
\vspace{3mm}
("Moreover I Advise a Super-Flavour Factory has to be Built")
}}
\footnote{Lecture given at ARGUS-Fest, Nov. 9, 2007, DESY, Hamburg, Germany}
\vspace*{8mm}

{\Large{\bf I.I.~Bigi}} \\
\vspace{2mm}

{\sl Department of Physics, University of Notre Dame du Lac}
\vspace*{-.8mm}\\
{\sl Notre Dame, IN 46556, USA}\\
{\sl email: ibigi@nd.edu}

\vspace*{5mm}

{\bf Abstract}\vspace*{-1.5mm}\\
\end{center}
The discovery of $B_d - \overline B_d$ oscillations twenty years ago by the ARGUS collaboration 
marked a watershed event. It persuaded a significant part of the HEP community that the large time 
dependent \cp~asymmetries predicted for some $B_d$ decays might be within the 
reach of specially designed experiments. This opened the successful era of the $B$ factories, which 
has a great future still ahead. After sketching the status of heavy flavour physics I describe why we need  to continue a comprehensive heavy flavour program not only for its intrinsic reasons -- 
it is even mandated as an integral part of the LHC program. Notwithstanding the great success anticipated for the LHCb experiment I explain why 
a Super-Flavour Factory is an essential complement to the LHC program. 

\noindent

\tableofcontents

\pagebreak


\noindent 
{\bf Prologue}

\vspace{0.3cm}

Earlier this afternoon we heard from Prof. Schopper how on his first visit here his request to be 
taken to DESY was misconstrued by the taxi driver. My experience this time was fundamentally different: 
when I told my taxi driver in Altona that I have to go to DESY, he immediately understood the nature of my destination. He perked up and said: "Oh, I am just reading a book on quantum chemistry -- can we talk about it?" I take my experience as re-assuring 
evidence for a growing appreciation of scientific culture. Yet the reality-based among you -- i.e. the 
experimentalists -- will probably think: "Typical theorist!" For looking at me you will realize that I 
am much older now than Prof. Schopper was {\em then}: therefore I -- unlike him -- was above suspicion. 

Allow me another brief look back. When I was invited before 1987 to give a talk and I suggested my topic -- you can easily guess, what it was \cite{BS80} -- I heard the following reaction: " Yes, yes, we know, Ikaros ..., but could you not talk about something relevant?" After ARGUS' discovery of 
$B_d - \overline B_d$ oscillations twenty years ago \cite{ARGUS87}, I never heard that again. Tony Sanda and I benefitted more from this discovery than most high energy physicists, and I can state an  emphatic:  
"Thank you, thank you, ARGUS!" 

At the time of ARGUS' discovery $B_d$ oscillations 
had been expected to proceed rather slowly.  The main reason for that prediction was that the UA1 experiment had reported strong evidence 
for having discovered top quarks with a mass of $40 \pm 10$ GeV. Almost all theorists accepted those findings. Peter Zerwas, however, did not, and he explained the reasons for his skepticism to me at the time. I should have listened to Peter -- it is the only time I did not, and I have been kicking myself for it ever since! 

\begin{figure}[ht]
\begin{center}
\epsfig{bbllx=0.5cm,bblly=16cm,bburx=20cm,bbury=23cm,
height=5truecm, width=12truecm,
        figure=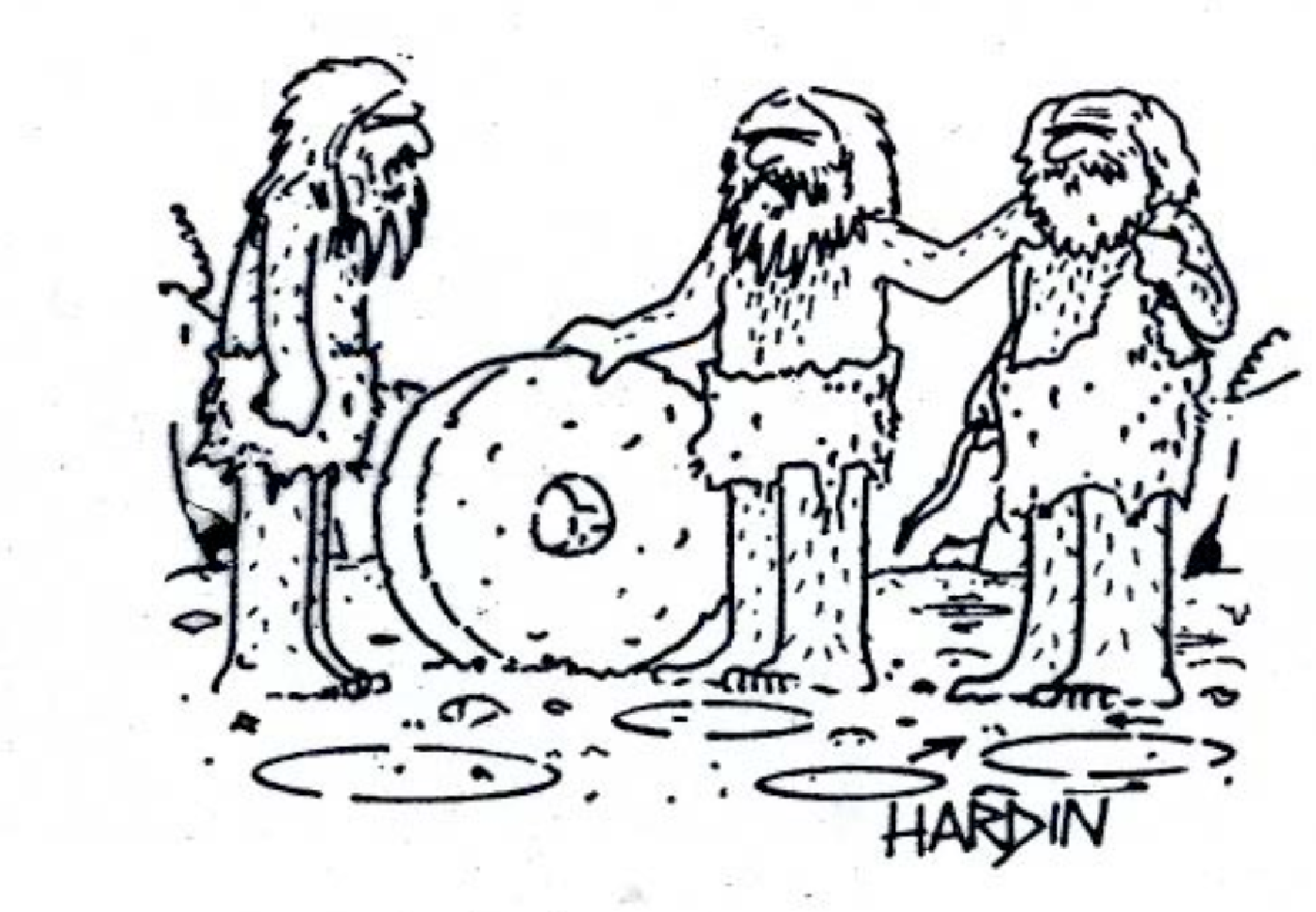}
\caption{" To be honest, I never would have invented the wheel if not for Urg's groundbreaking theoretical work with the circle." 
\label{WHEEL}  
}
\end{center}
\end{figure}

Our knowledge of $B$ meson dynamics has been expanded greatly over the last twenty years in a process accelerated by the success of the $B$ factories. This development has been helped by 
theorists in a way nicely expressed by the cartoon of Fig.\ref{WHEEL}, which I found last spring reading the In-flight journal of United Airlines: The chap in the middle, obviously an experimentalist, graciously 
-- if with a slightly patronizing flavour -- gives some credit to the theorist on his left by declaring: " To be honest, I never would have invented the wheel if not for Urg's groundbreaking theoretical work with the circle." 

I have given the first title of my talk in Latin based on a fundamental Catholic tenet recently re-confirmed by the new church leadership: 
{\em If it can be expressed in Latin, it must be true.} Since Hamburg is not exactly a hotbed of Catholicism, I will use a less august language, while fully aware that the elegance and cogency of the argument will suffer from this drawback. 

The talk will be organized as follows: In Act I I will sketch the role and status of studies of flavour dynamics; in Act II I will gaze into my crystal ball concerning the future of flavour physics as carried 
out for certain by LHCb and hopefully Super-Flavour Factories; in Act III I will present my conclusions before finishing with an Epilogue. 

\section{Act I -- On the Role and Status of Flavour Physics}
\label{ACT1}

Allow me to go "medias in res" rather than beat around the bushes. While the detailed study of 
strangeness changing processes was instrumental for the creation of the Standard Model (SM), that of 
charm changing ones was central for its acceptance, and that of 
beauty changing ones has almost completed the SM's validation (with only the Higgs boson not having been 
discovered yet). 

\begin{figure}[ht]
\begin{center}
\epsfig{
height=7truecm, width=7truecm,
        figure=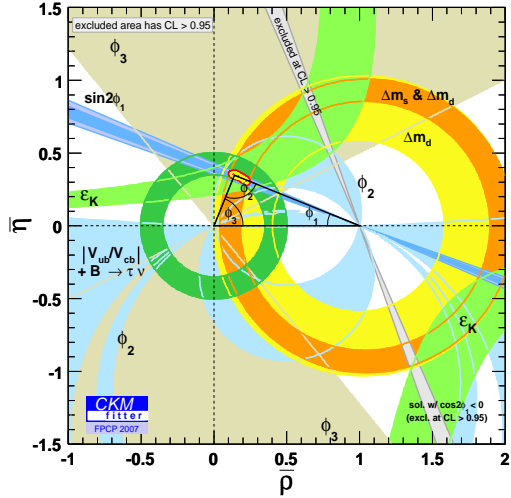}
\caption{The CKM Unitarity Triangle fit (courtesy CKM fitter collab.).
\label{CKMTRIANGLEFIT}  
}
\end{center}
\end{figure}

As explained in previous talks 
\cite{LIGETI,SCHUBERT}, the unitarity of the $3\times 3$ CKM matrix $V_{CKM}$ implies among others the following  relation among its (complex) elements:  
\beq
V^*_{ub}V_{ud} + V^*_{cb}V_{cd} + V^*_{tb}V_{td} = 0 \; , 
\label{CKMTRIANGLEEQ}
\eeq
which can be represented as a triangle in the complex plane. It is usually referred to as 
`the' CKM unitarity triangle. While the sides of the triangle reflect transition rates for $K$ and $B$ mesons (including pure quantum effects like oscillations), the angles determine \cp~asymmetries. 
Accordingly the area of the triangle is a measure for those asymmetries. Since 
re-scaling the triangle leaves the angles unchanged, one conveniently normalizes the base line 
to unit length. 
Our knowledge of flavour dynamics is sketched in a highly condensed 
form in Fig.\ref{CKMTRIANGLEFIT} by showing constraints from data -- most importantly from 
$\Delta M_{B_d}$, $\Delta M_{B_s}$ \cite{CDFBSPUB}, 
$|V_{ub}/V_{cb}|$ \cite{HFAG} and the \cp~sensitive observables $\epsilon_K$ and 
$\phi_1$ (a.k.a. $\beta$). The latter is the angle extracted from the time dependent \cp~asymmetry in $B_d \to \psi K_S$.  These constraints are inferred from a very heterogeneous set of transitions 
occurring on vastly different time scales. Yet they do overlap in a smallish domain indicated by the two ellipses for the apex of the triangle -- a highly non-trivial success for the SM! 

\begin{figure}[ht]
\begin{center}
\epsfig{bbllx=0.5cm,bblly=18cm,bburx=20cm,bbury=23cm,
height=5truecm, width=18truecm,
        figure=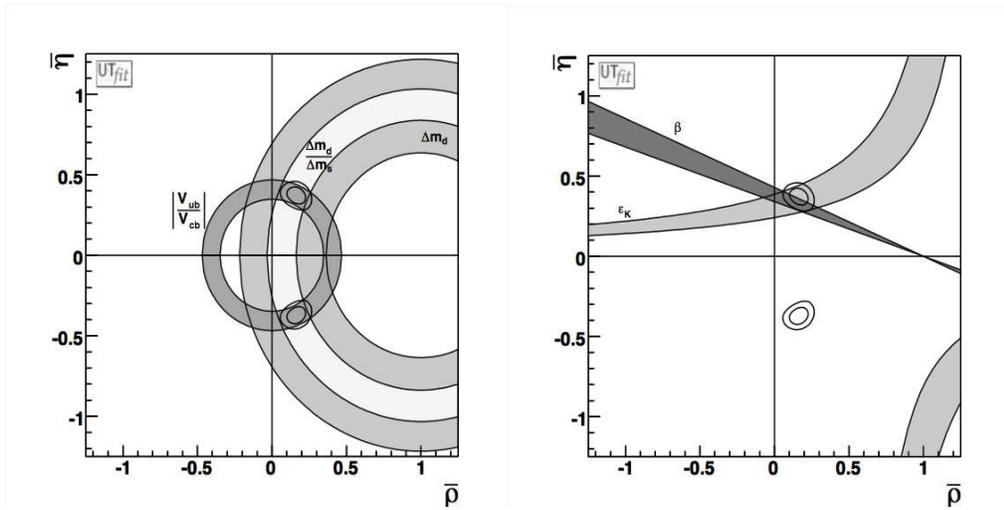}
\caption{CKM unitarity triangle from $|V(ub)/V(cb)|$ and $\Delta M_{B_d}/\Delta M_{B_s}$ on the left and  
       compared to constraints from 
$\epsilon_K$ and sin2$\phi_1$ on the right (courtesy V. Sordini). 
\label{PIERINI}  
}
\end{center}
\end{figure}

Fig.\ref{CKMTRIANGLEFIT} containing all constraints is very busy and thus obscures some of the relevant 
findings. Let me illuminate this by a highly topical example, namely the 
profound impact resolving $B_s - \bar B_s$ oscillations has had. Look at the left plot in Fig.\ref{PIERINI}. 
The triangle there is constructed from its three sides: the unit length baseline, and the other two sides as inferred from $|V_{ub}/V_{cb}|$ \cite{HFAG} and  
$\Delta M_{B_d}/ \Delta M_{B_s}$ \cite{CDFBSPUB}, respectively, with 
the widths of the bands denoting the uncertainties (mainly of a theoretical nature). The two bands overlap in a small domain, where the apex has to lie. The resulting triangle clearly has a non-zero area:  
from two \cp~insensitive observables -- i.e., two quantities that can be non-zero, even 
when \cp~invariance holds -- we can thus infer that the SM has to contain \cp~violation. Yet the situation is even more intriguing, as the right plot in Fig.\ref{PIERINI} shows: the amount of \cp~violation inferred from 
$|V_{ub}/V_{cb}|$ and  $\Delta M_{B_d}/ \Delta M_{B_s}$ is completely consistent with the observed 
\cp~asymmetries as expressed through $\epsilon_K$ and $\phi_1$ 
(a.k.a. $\beta$)! This marks another 
triumph for KM theory: From the observed values of two \cp~{\em in}sensitive observables one infers the size of \cp~asymmetries in even quantitative agreement with the data. 

So why not declare victory and close (the heavy flavour) shop? There are two sets of reasons against it: 
\begin{enumerate} 
\item 
We have experimental evidence of mostly heavenly origin that the SM is incomplete: neutrino 
oscillations, dark matter and dark energy. 
\item 
The novel successes the SM has scored 
since the turn of the millenium -- having the predictions of truly large \cp~asymmetries in $B$ 
decays confirmed -- do not illuminate any of its mysterious features; if anything, they deepen the mysteries: 
\begin{enumerate}
\item 
Theoretical arguments centered on the `gauge hierarchy problem' strongly suggest that the electroweak symmetry breaking is driven by something beyond the SM's $SU(2)_L\times U(1)$ gauge theory 
with that something entering around the TeV energy scale. 
Those arguments have been sufficiently persuasive as to motivate the construction of the LHC complex at CERN, and I will refer to it as the "confidently predicted New Physics" ({\em cpNP}). A popular candidate is provided by SUSY. 
\item 
We have no structural explanation for charge quantization and the lepton-quark connection; 
i.e., why is the electric charge of the electron exactly three times that for $d$ quarks? A natural resolution of this puzzle arises in Grand Unified Theories, which place quarks and 
leptons into the same multiplets. I will refer to it as the "guaranteed New Physics" ({\em gNP}) characterized 
by scales of the order of about $10^{14}$ GeV; an $SO(10)$ gauge theory provides an attractive scenario. 
\item 
It seems likely that family replication and the hierarchical pattern in the CKM parameters is created by some fundamental dynamics operating at some high scale. I will call it "strongly suspected 
New Physics" ({\em ssNP}). We do not know what that scale is, and expressing the hope that M theory will resolve this puzzle is a polite way of saying that we have hardly a clue about it. 
\end{enumerate}
Detailed and comprehensive heavy flavour studies might -- just might -- provide insights into the 
{\em gNP} and {\em ssNP} -- i.e., items (b) and (c) above -- although we cannot count on it. Yet they are likely to be essential for identifying the {\em cpNP}, item (a)! 
\end{enumerate}
Let me explain the last point in some detail:
\begin{itemize}
\item 
I am confident the LHC will reveal the presence of New Physics directly by the production of new quanta. 
\item 
Yet we should aim higher than `merely' establishing the existence of such New Physics. The goal must be to identify its salient features. I am a big fan of SUSY, yet we should remember that SUSY per se is not a theory or even class of theories -- it is an organizing principle. 
\item 
TeV scale dynamics is likely to have some impact on $B$, $D$ and $\tau$ decays. We need to probe the discovery potential in those processes in order to identify the New Physics. 
{\em A dedicated heavy flavour program is not a luxury -- it is integral to the core mission of the LHC program}. 
\item 
We should already have seen, say, the impact of a `generic SUSY' \cite{MURAGEN} 
-- i.e.,  a version of SUSY picked at random out of the multitude of SUSY implementations. On the other hand past experience shows that Nature has not exhibited much taste for generic dynamics. Furthermore the one aspect of SUSY that is beyond dispute, namely that it is broken, is also the least understood one. 
\item 
The often heard term of `minimal flavour violation' is a classification scheme 
\cite{MFV02}, not a theory -- analogous to the case of the `superweak model' of \cp~violation. We have to ask to which degree do dynamics implement such a scenario: does it represent a strict or -- more likely -- an approximate one? 

\end{itemize}
To summarize: we need to continue a comprehensive program of experimental heavy flavour studies,  
not to shed light on the flavour mystery of the SM -- although that might happen -- but as a high sensitivity instrument for probing 
more fully the dynamics behind the electroweak phase transition. We have learnt (and some of us had actually predicted it several years ago \cite{CPBOOK}) that heavy flavour transitions typically 
will {\em not} be affected in a {\em numerically massive} fashion by the anticipated New Physics. Yet this should make us strive for higher sensitivity in our searches, not to abandon them. 

\section{Act II: On the Future -- LHCb and Super-Flavour Factories}
\label{Act2}

Looking at the next few years I am pleased to say that the state of heavy flavour studies is promising and strong. The contributions from the CDF and D0 experiments studying hadronic collisions have greatly exceeded expectations with respect to $B$ physics. The latest example -- and a spectacular one -- was the measurement of $B_s - \bar B_s$ oscillations \cite{CDFBSPUB}. More than a decade ago LHCb with its focus on $B$ physics was approved as an experiment to take data from day one of LHC's operation. 
The European HEP community deserves credit for this visionary decision. 
I am confident that LHCb will make truly seminal contributions in particular in the exploration of $B_s$ decays -- most notably the time dependent \cp~asymmetries in $B_s \to \psi \phi$, $\phi \phi$. 
Since $B_d$ and $B_s$ transitions a priori represent different chapters in Nature's book of dynamics, 
we better analyze both with high accuracy. 
There is no doubt in my mind that the HEP community will reap great benefits from the support it gives to LHCb. 

\subsection{The "Second Renaissance" of Charm Physics}
\label{CHARMREN}

The case for a continuing experimental program of heavy flavour physics has been strengthened considerably by the strong evidence presented by Belle and BaBar in the spring of 
2007 \cite{BELLEOSC1,BABAROSC,BELLEOSC2}. Analogous 
to the $B_d$ case $D^0 - \bar D^0$ oscillation rates can be expressed in terms of the calibrated mass and width differences between the two mass eigenstates: 
$x_D \equiv \Delta M_D/\overline \Gamma _D$, $y_D \equiv \Delta \Gamma_D/2\overline \Gamma _D$. 
Averaging over all relevant data -- an intriguing enterprise, yet one that is not without risk at present -- 
one obtains \cite{HFAG}
\beq 
x_D = (0.87 ^{+0.30}_{-0.34}) \cdot 10^{-2} \; \; , \; \; y_D = (0.66 ^{+0.21}_{-0.20}) \cdot 10^{-2} \; , 
\eeq
which represents 5 $\sigma$ evidence for $(x_D,y_D) \neq (0,0)$. 

If we had observed $x_D > 1\% \gg y_D$, we would have a strong prima facie case for New Physics -- but such a scenario has been basically ruled out now. For the data point to 
$x_D \sim y_D \sim 0.5 - 1\%$. 
\begin{enumerate}
\item 
Effects of that size could be due `merely' to SM dynamics \cite{BUDOSC,FALK1}. Even then it would be a seminal discovery and should be measured accurately; for it can help to validate the observation of time dependent \cp~asymmetries as discussed below. 
\item 
At the same time $D^0 - \bar D^0$ oscillations can still receive sizable contributions from New Physics. 

\end{enumerate}
How can we resolve this conundrum? 
\begin{itemize}
\item
We might be just one theoretical breakthrough away from a more accurate SM prediction. Maybe. 
\item 
Rather than wait for that to happen, since it might take a while, the experimentalists might follow the Calvinist tradition of demonstrating heavenly favour by achieving earthly success. For they can search  
for \cp~violation in charm transitions. It is most appropriate to emphasize this option at this ARGUS-Fest. Will history repeat itself in the sense that the discovery of oscillations will 
prompt a program  of \cp~studies? There are obvious challenges involved: We are dealing with a 
`centi-ARGUS' scenario, since $x_D$ is about a factor of hundred smaller than $x_{B_d}$. I think 
our experimental colleagues will learn to deal with that. Another difference is that KM theory does not predict sizable, let alone large effects in the charm system. I submit this is actually an advantage, since the ratio of signal to `theoretical noise' (from SM contributions) might well be large. Furthermore we are not engaging in a `wild goose chase' here, since baryogenesis requires New Physics with \cp~violation. 
\end{itemize}
The decay channels being analyzed for oscillations \cite{CICERONE} -- 
$D^0 \to K^+K^-/\pi^+\pi^-/K_S\pi^+\pi^-$ --  
are also excellent targets for such searches. For oscillations can generate {\em time dependent}  
\cp~asymmetries there. No such effects have been seen so far -- but the experimental sensitivity has only recently reached a domain, where one could hope for a signal \cite{GKN,CHINAREV}. Consider 
\beq 
D^0 \to K^+K^-
\eeq
In qualitative analogy to $B_d \to \psi K_S$ the oscillation induced \cp~asymmetry is given by 
\beq 
\frac{{\rm rate}(D^0 (t) \to K^+K^-) - {\rm rate}(\overline D^0 (t) \to K^+K^-)}
{{\rm rate}(D^0 (t) \to K^+K^-) + {\rm rate}(\overline D^0 (t) \to K^+K^-)} \sim 
x_D\;  [{\rm or}\; y_D] \cdot \frac{t}{\tau_D} \cdot {\rm sin} \phi_{\rm weak}  \; ; 
\eeq
i.e., it is by and large bounded by the value of $x_D$ [or $y_D$]. If those do not exceed the 1\% level, 
nor can the asymmetry, and that is about the experimental sensitivity at present. Having seen a 
signal would hardly have been credible. Yet now it is getting interesting; for any improvement in experimental sensitivity might reveal an effect.

\subsection{The Case for a Super-Flavour Factory}
\label{SFF}

I count on LHCb to become a highly successful experiment in heavy flavour studies -- 
benchmark transitions like $B_s \to \psi \phi$, $\phi \phi$ or $D^0 \to K^+K^-$, $K^+\pi^-$ 
are optimal for LHCb's consumption -- yet it will {\em not} complete the program! 

As indicated above we can typically expect at most moderate deviations from SM predictions. 
Precision is therefore required both on the experimental and the theoretical side. The latter 
requires `flanking measures'; i.e., in order to calibrate our theoretical tools for interpreting decay 
rates, we want to analyze final states with (multi)neutral hadrons like 
$B^0 \to \pi^+\pi^-\pi^0/3\pi^0$, $B^- \to \pi^-\pi^0\pi^0$. We need to study 
$B_d \to \phi K_S$, $\eta^{(\prime)}K_S$ with precision, since those lessons are complementary 
rather than repetitive to those inferred from $B_s \to \phi \phi$. {\em In}clusive reactions can be 
described more reliably than {\em ex}clusive ones -- a valuable asset when searching for smallish 
effects. We want to measure also semileptonic $B$ decays -- $B\to \tau \nu D/\tau \nu X$ -- as a probe 
for the exchange of charged Higgs bosons with a mass in the several hundred GeV range. Comprehensive \cp~studies in charm transitions are mandated now more than ever before due to the 
strong evidence for $D^0 - \overline D^0$ oscillations. Last, but most certainly not least we have to search 
for both lepton flavour and \cp~violation in $\tau$ decays.  

A Super-Flavour Factory -- a low-energy $e^+e^-$ machine with a luminosity of 
$10^{36}$ cm$^{-2}$ s$^{-1}$ is needed to take on these challenges \cite{BROWDER}. In this context let me express a warning: a Super-Flavour Factory requires a very different kind of justification than the original 
$B$ factories at KEK and SLAC did. For those we had so-called `killer applications' 
\cite{BS80}; i.e., effects that {\em individually} would have an immediate and profound impact on the SM, if they were observed or ruled out. Those were the time dependent \cp~asymmetries in $B_d \to \psi K_S/\pi^+\pi^-$; for they were predicted -- 
{\em with no plausible deniability} -- to reach the several $\times$ 10 \% range; this was inferred from the 
only known \cp~violation in the early 1990's, namely $K_L \to \pi \pi$, which is characterized by  
$|\epsilon_K| \simeq 0.22\%$. Furthermore the domain of quantitative heavy flavour dynamics was still largely `virgin' territory. The success of the $B$ factories has greatly exceeded our expectations: they have promoted the KM paradigm from an ansatz to a {\em tested theory}. As far as \cp~violation in the decays of hadrons is concerned, we no longer look for alternatives to KM theory, only to corrections to it.  However, the very success of the $B$ factories has raised the bar for a Super-Flavour Factory. Rather than exploring unchartered territory, we want to revisit it, albeit with greatly enhanced sensitivity. 
It is like going back into a heavily mined gold mine. 

To say it slightly differently. There are two types of research programs, namely `hypothesis driven' 
and `hypothesis generating' research. While the former tests an existing paradigm (and thus is 
favoured by funding agencies), the latter aims at developing a new paradigm.  
The program at the $B$ factories belonged to the former variety -- and represents a most successful 
one -- yet a Super-Flavour Factory aims at the latter by searching mainly for the 
anticipated `New \cp~Paradigm'. 

The top priority at a Super-Flavour Factory has to be assigned to studies of $B$ physics, which still has a rich agenda as explained in the talks by Ligeti \cite{LIGETI} and 
Golutvin \cite{GOLUTVIN}; for more details see Ref.\cite{BROWDER}. I will not repeat their discussion here and 
instead sketch the agenda of two other areas accessible at a Super-Flavour Factory, namely charm and $\tau$ physics. 

\subsubsection{2nd Priority: \cp~Studies in Charm Transitions}

I had mentioned before that the observed rate of $B_s - \overline B_s$ oscillations is consistent with 
the SM prediction within the latter's significant uncertainty. The potential New Physics hiding behind 
the uncertainty can be revealed in the time dependent \cp~asymmetry in $B_s \to \psi \phi$, since the latter is small in the SM for reasons germane to it \cite{BS80}. 

The same strategy can and should be pursued 
in charm transitions. While the observed oscillation rate is not clearly inconsistent with the SM, the 
uncertainties are quite large.  Yet decisive tests can be provided by \cp~studies in 
$D^0 \to K^+K^-/\pi^+\pi^-/K^+\pi^-/K_S\pi^+\pi^-$ as mentioned before, since the  
`signal to theoretical noise' ratio is very likely higher in \cp~asymmetries than in pure oscillation phenomena.  
For the former are shaped to a higher degree by short-distance dynamics, over which we have better theoretical control than over the non-perturbative long-distance dynamics. Furthermore 
KM theory allows for only small asymmetries to arise in a rather restricted set of channels 
\cite{CICERONE}. 

I want to add two examples of a bit unorthodox nature. 

{\bf The `Dark Horse': Semileptonic $D^0$ Decays}

\noindent 
In analogy to the $B_d$ case, the emergence of `wrong-sign' leptons -- 
$D^0 \to l^-\overline \nu K^+$ or  $\overline D^0 \to l^+ \nu K^-$ -- signals 
oscillations have taken place. We already know that unlike for $B_d$ mesons it is a rare 
process for neutral charm mesons. Once we have accumulated such wrong-sign events, we can ask whether this rate 
is different for the meson and anti-meson transition: 
\beq 
a_{SL}(D^0) \equiv \frac{\Gamma (D^0 \to l^-\overline \nu K^+) - \Gamma (\overline D^0 \to l^+ \nu K^-)}
{\Gamma (D^0 \to l^-\overline \nu K^+) + \Gamma (\overline D^0 \to l^+ \nu K^-)}
\eeq
Such differences have been and are being searched for in the semileptonic decays of neutral $K$ and $B$ mesons. For $K_L$ decays the expected rate has been found -- 
$a_{SL}(K_L)\simeq 3.3 \cdot 10^{-3}$; the experimental upper bounds for neutral $B$ mesons 
have not yet reached the SM predictions: $a_{SL}(B_d)\simeq 4 \cdot 10^{-4}$, 
$a_{SL}(B_s)\simeq 2 \cdot 10^{-5}$ \cite{LENZ}. 
We understand why these numbers are so tiny. For $a_{SL}$ is given very roughly by 
\beq
a_{SL} \sim \frac{\Delta \Gamma}{\Delta M} \cdot {\rm sin}\phi_{\rm weak} \; . 
\eeq
While $\Delta \Gamma/\Delta M \simeq 1$ for kaons, we have sin$\phi_{\rm weak}$ $\ll 1$ due to the 
third quark family being almost decoupled from the first two. For $B_d$ it is the other way around: 
$\Delta \Gamma/\Delta M \ll 1$, yet sin$\phi_{\rm weak} \sim {\cal O}(0.1)$. For $B_s$ mesons we have furthermore sin$\phi_{\rm weak}$ $\ll 1$, since on the leading level only the second and third quark 
family contribute. 

A rough estimate yields $a_{SL}(D^0)|_{SM} \leq 10^{-3}$. Present data suggest 
$\Delta \Gamma/\Delta M$ to be about unity. With 
New Physics inducing a weak phase we could conceivably obtain a relatively large value:  
$a_{SL}(D^0) \sim \; {\rm few} \times 10^{-2}$; i.e., while we know that semileptonic $D^0$ decays 
produce few wrong-sign leptons, they might exhibit a large \cp~asymmetry -- in marked contrast to 
$K_L$, $B_d$ and $B_s$ mesons. 

{\bf Final State Distributions, \ot~odd Moments}

\noindent 
So far all \cp~violation has been found in partial widths -- except for one, the forward-backward 
asymmetry in the orientation of the $\pi^+\pi^-$ and $e^+e^-$ planes in $K_L \to \pi^+\pi^-e^+e^-$. 
It had been predicted \cite{SEGHALKL} and subsequently found that the expectation value for this angular asymmetry 
is about 14\% \cite{PDG06} -- yet driven by $|\epsilon_K| \simeq 0.23\%$. How can that be? This  
puzzle is resolved, when one realizes that both amplitudes that generate the asymmetry through their 
interference -- $K_L \stackrel{CPV} \to \pi^+\pi^- 
\stackrel{E1}\to \pi^+\pi^- \gamma^* \to \pi^+\pi^- e^+e^-$ and 
$K_L \stackrel{M1}\to \pi^+\pi^- \gamma^* \to \pi^+\pi^- e^+e^-$ -- 
are greatly suppressed, albeit for different reasons: it is the \cp~violation in the first and the $M1$ 
feature in the second amplitude.  Such a dramatic enhancement of the 
asymmetry does not come for free, of course: 
the price one pays is a tiny branching ratio of about $3\cdot 10^{-7}$; i.e., one trades 
branching ratio for size of the asymmetry. This is a very desirable trade -- if one has a copious 
production source. 

There might be a close analogy in the charm complex, namely in the angular 
distribution of the $K^+K^-$ relative to the $\mu^+\mu^-$ plane in 
\beq 
D_L \to K^+K^- \mu^+\mu^-  \; , 
\eeq
where a \cp~violating $E1$ amplitude interferes with a \cp~conserving $M1$ amplitude to generate 
a forward-backward asymmetry. The latter could exhibit an enhancement of the underlying 
\cp~violation leading to $D_L \to K^+K^-$ by an order of magnitude depending on details of the strong 
dynamics. 
This radiative decay has not been observed yet; its branching ratio 
could be as `large' as about $10^{-6}$. 

The reader might view this discussion as completely academic, since it requires a pure sample of 
long-lived neutral $D$ mesons in qualitative analogy to $K_L$. Yet since the lifetime difference 
between $D_L$ and $D_S$ can hardly reach even the 1\% level, `patience' -- waiting for the 
$D_S$ component to decay away -- is insufficient. Yet there is a unique capability of a Super-Flavour Factory that can be harnessed here through the use of EPR correlations \cite{EPR} or `entanglement'. 
Consider running at charm production threshold: 
\beq 
e^+e^- \to \psi ^{\prime \prime} (3770) \to D_S D_L  \; . 
\eeq
Once one of the neutral $D$ mesons decays as $D \to K^+K^-$, we know unambiguously that the other meson has to be a $D_L$, as long as \cp~is conserved. We can then track its decays into the 
$K^+K^-\mu^+\mu^-$ final state.

\subsubsection{3rd Priority: $\tau$ Physics}

{\bf Lepton Flavour Violating Decays (LFV)}

\noindent 
Finding a transition of the type $\tau \to l \gamma$ or $\tau \to 3 l$ establishes the existence of New Physics, since lepton flavour is violated. The $B$ factories have established upper bounds of 
few$\times 10^{-8}$. The range $10^{-8} - 10^{-10}$ is a very promising search domain rather than an ad hoc one. For several classes of New Physics scenarios -- in particular of the GUT variety with their 
connections to $\mu \to e \gamma/3e$ -- point to that range \cite{BROWDER}. The radiative transition 
$\tau \to l \gamma$ seems to be clearly beyond the reach of LHC experiments; this might well turn out to be true for $\tau \to 3 l$ as well. Yet a Super-Flavour factory can push into this domain and possible sweep it out. 

{\bf \cp~Violation in $\tau$ Physics}

\noindent 
The next great challenge in \cp~studies is to find \cp~violation in leptodynamics. The leading contenders are the electron EDM, \cp~asymmetries in neutrino oscillations and in semi-hadronic $\tau $ decays like $\tau \to K\pi (\pi)\nu$ \cite{KUEHN,BSTAU}. If found, it would `de-mystify' \cp~violation as a phenomenon 
present both in the quark and lepton sectors. Maybe more importantly it would provide us with a potential benchmark for leptogenesis that can subsequently induce baryogenesis in our Universe. There will not be any competition from LHC experiments for probing \cp~symmetry in $\tau$ decays. At a  
Super-Flavour Factory one can also employ a unique and powerful tool, namely longitudinal beam polarization: it will lead to the production of polarized $\tau$ leptons, which provides another handle on 
\cp~invariance \cite{TSAI,BSTAU}. 

\noindent 
For proper perspective one should note that while a LFV rate has to be {\em quadratic} in a 
New Physics amplitude, a \cp~asymmetry (in a SM mode) is linear only: 
\beq 
\cp~{\rm odd} \; \sim \; |T^*_{\rm SM}T_{\rm NP}|   \; \; \; {\rm vs.}\; \; \; 
{\rm LFV} \; \sim \; |T_{\rm NP}|^2 \; . 
\eeq
Observing a $10^{-3}$ [$10^{-4}$] \cp~asymmetry in 
$\tau \to K\pi\nu$ then corresponds very roughly to discovering $\tau \to \mu \gamma$ with a branching  ratio of 
about $10^{-8}$ [$10^{-10}$].

\subsection{Design Criteria for a Super-Flavour Factory}

The preceding discussion leads to the following strategic goals when designing a 
Super-Flavour Factory: 

\noindent 
$\bullet$ {\em You cannot overdesign a Super-Flavour Factory}. If what we know now about the 
size of the \cp~asymmetry in $B_d \to \psi K_S$ had been known when the $B$ factories were 
proposed, a less ambitious target for the luminosity would most likely have been chosen. In retrospect 
both $B$ factories had been over-designed -- yet that is exactly what was a cornerstone of their 
spectacular success! What is true for a `hypothesis driven' research program, is even more true for a 
`hypothesis generating' one. Tony Sanda's dictum "We need a luminosity of 
$10^{43}$ cm$^{-2}$ s$^{-1}$" is certainly `tongue-in-cheek', but not frivolous in that sense. If you must 
stage the construction, do {\em not} compromise on final performance. To be more down to earth: a 
data sample of 10 ab$^{-1}$ -- an increase by an order of magnitude over the existing set -- should be targeted as an intermediate step; in the end one should aim for at least 50 ab$^{-1}$. 

\noindent 
$\bullet$ Keep the background as low as possible. 

\noindent 
$\bullet$ Make the detector as {\em hermetic} as possible. This is essential when aiming for 
$B \to \nu \bar \nu K^{(*)}...$, $B \to \tau \nu D...$, $D_{(s)} \to \tau \nu$ modes. 

\noindent 
$\bullet$ Keep the flexibility to eventually have quality runs on the $\Upsilon (5S)$ 
resonance, be it for calibrating {\em absolute} rates for $B_s$ transitions or analyzing some of their features that could not be settled by LHCb. 

\noindent 
$\bullet$ It might turn out to be even more important to be able to run in the charm threshold region 
with good luminosity to reduce systematic uncertainties when searching for tiny \cp~asymmetries in charm decays. For the background is lowest there; furthermore quantum correlations can be harnessed to obtain unique information \cite{CICERONE}. I have mentioned just one example, namely the ability to prepare a `beam' of $D_L$ mesons. 

\noindent 
$\bullet$ Make a reasonably strong effort to obtain at least one longitudinally polarized beam. 
This is an essential tool in probing \cp~invariance in the production and decay of 
$\tau$ leptons. It would also be valuable in dealing with the background when searching 
for LFV $\tau$ decays (and for some \cp~asymmetries in charm baryon decays). 

\section{Conclusions and Outlook}

We are about to embark on a most exciting adventure: we stand at the beginning of an era that promises  
to reveal the dynamics behind electroweak symmetry breaking.  The 
central stage for this adventure will be the LHC, where quanta signaling New Physics are expected to 
be produced. Since failure of the LHC program would have disastrous consequences for the 
future of fundamental physics, it just cannot be tolerated!  
Yet heavy flavour studies probing the family structure and 
\cp~symmetry in the $K$, $D$, $B$ and $\tau$ sectors will be central players in the evolving drama. 
\begin{itemize}
\item 
Such studies are and will remain of fundamental importance in our efforts of revealing `Nature's Grand 
Design'; 
\item 
their lessons cannot be obtained any other way; 
\item 
they cannot become obsolete. 

\end{itemize} 
At the same time comprehensive studies of \cp~violation, oscillations and rare decays 
can be {\em instrumentalized} to analyze the anticipated TeV scale New Physics. I see three scenarios play out over the next several years: 
\begin{enumerate}
\item 
The `optimal' one: New Physics has been discovered in high $p_{\perp}$ collisions at the LHC. 
Then we must determine its salient features, and this cannot be done without analyzing its 
impact on flavour dynamics -- even if there is none! With the mass scale of the New Physics 
revealed directly, lessons from heavy flavour rates can be interpreted with more 
quantitative rigour. 
\item 
The `intriguing' one: deviations from SM predictions have been established in heavy flavour decays. 
\item
The `frustrating' one: no deviations from SM predictions have been identified anywhere. 

\end{enumerate}
I bet it will be the first scenario with some elements of the second one. We should not overlook that heavy flavour studies can realistically have sensitivities up to the about 10 - 100 TeV scale -- 
well beyond the direct reach of the LHC. But in any case none of these scenarios weaken the essential role of flavour studies. For even the `frustrating' 
scenario does not resolve any of the central mysteries of the SM.  
\footnote{This is of course a purely scientific-intellectual argument -- the political one would play out very 
differently.} 

The LHCb experiment will be a worthy and successful standard bearer of heavy flavour physics, yet 
it will not complete the program. The era of the heavy flavour factories inaugurated by ARGUS' 
discovery twenty years ago has not run its profitable course yet -- the best might actually still be ahead. 
A Super-Flavour Factory provides unique capabilities in searching for LFV and \cp~violation in 
$\tau$ decays, unmatched access to \cp~studies in charm transitions and measurements of $B$ 
decays that are highly complementary to the LHCb program. The HEP community is fortunate to have a battle tested and enthusiastic `army' to embark on a Super-Flavour Factory campaign and will benefit greatly from the results of the latter.

\vspace{0.5cm}

\noindent 
{\bf Epilogue}

\vspace{0.3cm}

When we look back over the last thirty years -- i.e. including the period leading up to ARGUS' 
discovery of $B_d - \overline B_d$ oscillations -- we see several strands of developments: 
from the `heavy flavour sweatshops' -- ARGUS, CLEO and MARKIII -- to the present $B$ and 
tau-charm factories -- Belle, BaBar, CLEO-c and BESIII -- hopefully to a Super-Flavour Factory; 
accelerators pushing the high energy frontier -- the SPS, Tevatron, LEP I/II and SLC -- 
leading to the LHC and hopefully to the ILC; last (and presumably least for some of the readers) theory. 
These strands are not isolated from each other, but substantially intertwined. The generational challenge facing us is to understand the electroweak phase transition. This will be tackled in a dedicated way at the high energy frontier by the LHC experiments Atlas and CMS and at the high sensitivity 
frontier by LHCb. 
Yet they are unlikely to complete the task -- we will need more precise and more 
comprehensive data. This is where the ILC, which is also a top factory, and a Super-Flavour Factory come in as essential parts 
of the adventure. 

\begin{figure}[ht]
\begin{center}
\epsfig{bbllx=0.5cm,bblly=7cm,bburx=20cm,bbury=23cm,
height=6truecm, width=7truecm,
        figure=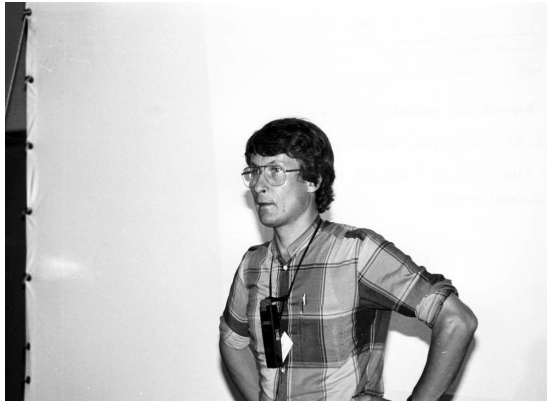}
\caption{Giving a talk in Heidelberg in 1986
\label{HD86}  
}
\end{center}
\end{figure}

\begin{figure}[ht]
\begin{center}
\epsfig{
height=4truecm, width=3truecm,
        figure=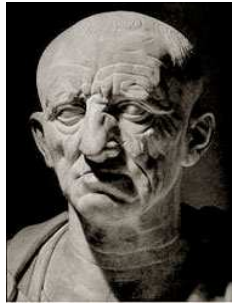}
\caption{Cato the Elder
\label{CATO}  
}
\end{center}
\end{figure}

Let me allow a very personal look back as well: Fig.\ref{HD86} shows me giving a talk at the Heidelberg Heavy Quark Symposium in 1986. Fig.\ref{CATO} on the other hand might be closer to how some see me now. It actually shows the person whose most famous quote I adapted for the title. 

It has been said: "All roads lead to Rome." Personally I think Rome is never a bad destination.  
When I said before we are at the beginning of an exciting journey into the unknown I was  incorrect, as shown by celebrating ARGUS' seminal achievements: For it is actually the continuation 
of an age-long adventure, and we are most privileged to be able to participate in it.

\vspace{0.5cm}

{\bf Acknowledgments:} This work was supported by the NSF under the grant number PHY-0355098. 
I am grateful for DESY for organizing and supporting  a very enjoyable ARGUS Fest, which gave me the 
opportunity to meet dear colleagues again, remember many things and learn about others for the first time.

\vspace{4mm}


\end{document}